\documentstyle[psfig,emulateapj]{article}

\input psfig.sty

\begin{document}

\title{Simulating Voids}

\author{David M. Goldberg \& Michael S. Vogeley\\
Department of Physics, Drexel University, Philadelphia, PA 19104}

\begin{abstract}
  We present a novel method for simulation of the interior of large
  cosmic voids, suitable for study of the formation and evolution of
  objects lying within such regions. Following Birkhoff's theorem,
  void regions dynamically evolve as universes with cosmological
  parameters that depend on the underdensity of the void. We derive
  the values of $\Omega_M$, $\Omega_{\Lambda}$, and $H_0$ that
  describe this evolution.  We examine how the growth rate of
  structure and scale factor in a void differ from the background
  universe. Together with a prescription for the power spectrum of
  fluctuations, these equations provide the initial conditions for
  running specialized void simulations. The increased efficiency of
  such simulations, in comparison with general-purpose simulations,
  allows an improvement of upwards of twenty in the mass resolution.
  As a sanity check, we run a moderate resolution simulation
  ($N=128^3$ particles) and confirm that the resulting mass function
  of void halos is consistent with other theoretical and numerical
  models.
\end{abstract}


\section{Introduction}

Large cosmic voids provide a unique testbed for models of galactic
formation and evolution (Peebles 2001). In recent years, there has
been intense effort to identify and study voids in a number of large
surveys such as the IRAS 1.2 Jy Survey (El-Ad et al. 1997), the PSCz
Redshift Survey (Plionis \& Basilakos 2002; Hoyle \& Vogeley 2002),
the Las Campanas Redshift Survey (Arbabi-Bigdoli \& M\"{u}ller 2002)
and the Updated Zwicky Catalog (Hoyle \& Vogeley 2002). Observed large
voids have typical diameters of $D\sim 25h^{-1}$Mpc, density contrast
$\delta\sim -0.9$, and fill at least 40\% of the universe (Hoyle \&
Vogeley 2002).  The properties of galaxies in voids clearly differ
from those in higher-density regions (Grogin \& Geller 1999a,b; Rojas
et al. 2003a,b). Recent studies using samples of $10^3$ void galaxies
from the Sloan Digital Sky Survey demonstrate that void galaxies are
bluer, fainter, more disklike, and have higher specific star formation
rates (Rojas et al. 2003a,b).

A variety of simulation methods has been applied to study formation
of object in voids, including application of Semi-Analytic Models
(SAMs) to compute properties of galaxies associated with dark matter
halos in cold dark matter simulations (Benson et al. 2003; Mathis \&
White 2002), simulations including hydrodynamics (Ostriker et al.
2003), and specialized high-resolution simulations of dark matter
halos in voids (Gottl\"ober et al. 2003).  Studying the
characteristics of galaxies in the context of large-scale numerical
simulations can be extremely costly because the region of interest
represents only $\sim 6-8\%$ of the total mass of the simulation, and
less than half the volume. Thus, enormous amounts of computing time
must be spent in order to study voids in a particular cosmological
model. Studies over many models naturally become even more difficult.

In principle, simulation of voids is far easier than simulation of a
typical large patch of the universe, because voids are dynamically
much simpler systems than clusters.  Large voids remain in the linear
regime for relatively longer time and approach spherical symmetry for
much of their evolution.  Birkhoff's theorem (Birkhoff 1923) tells us
that the internal dynamics of a spherically symmetric system will be
independent of the dynamics of the outside universe.  This theorem has
been applied in the development of spherically symmetric infall models
for galaxies (e.g., Gunn \& Gott 1972; Schechter 1980), cluster infall
regions (Reg\"os \& Geller 1989), for large-scale inhomogeneities in
general (Silk 1974), and for studying the population of Ly$\alpha$
clouds in voids (Manning 2002).  Voids can be approximated as
expanding, isolated universes unto themselves which do not accrete
matter from the universe at large.  As we describe below, this is only
an approximation, but one that is useful for studying formation of
objects well within the edges of voids.

We propose to exploit this simplicity to simulate universes entirely
composed of void-like regions.  A patch anywhere within such a
simulation would mimic the interior of a large void and allow
investigation of such issues as: What are galaxies like in such a
universe? At what epoch did they form?

The organization of this paper is as follows.  In
\S\ref{sec:parameters}, we derive the basic relations between the
cosmological parameters: expansion coefficients, and local values of
$\Omega_M$, $\Omega_\Lambda$ and $H$, in the void region and outside
of it.  In \S\ref{sec:powerspec}, we discuss the primordial power
spectrum which should be used in running the simulations.  
Details of applying this approach are discussed in \S\ref{sec:recipe}.
In \S\ref{sec:massfunc}, we discuss the results of a small simulation
meant as a ``sanity check'' on the proposed method.  
Finally, in
\S\ref{sec:future}, we discuss these results and future applications.

\section{Cosmological Parameters in the Voids}
\label{sec:parameters}

We can use Birkhoff's theorem to explore the physics of underdense
regions, because voids have the properties that they are largely
isolated throughout their evolution, and they are roughly spherical.
Birkhoff's theorem tells us that a spherically symmetric region with
specified mean density, cosmological constant density, radius $r$, and
expansion rate $\dot{r}$ will evolve exactly like a universe with a
Hubble constant $\dot{r}/r$, and a mean density equal to that of the
spherical region.  The growth of perturbations within that region will
exactly mimic (up to long-range tidal forces) structure growth in a
universe with the fiducial cosmology.

For the purpose of our discussion, we assume a background cosmology
with a pure CDM+Baryon matter density of $\rho_0$ at the present
(throughout, we will use the convention that a subscript ``0'' denotes
the value of the parameter at the present).  All calculations assume a
negligible density of relativistic matter.  The matter density
parameter may be expressed as $\Omega_{M}=8\pi G \rho_0/3 H_0^2$.  We
also assume a known cosmological constant $\Lambda$ with corresponding
density parameter $\Omega_{\Lambda}$.
   
At an arbitrarily high redshift, when the scale factor
$a_i=a(t_i)=1/(1+z_i)$, consider a spherical underdense region with
physical size $r_v(t_i)$ (henceforth, subscript ``v'' denotes a
property of the void region).  If the density contrast of the void is
$\delta_v(t_i)$, then the mass contained in the region is, of course,
\begin{equation}
M_v=\frac{4\pi}{3}r_v(t)^3[1+\delta_v(t)]\frac{\rho_0}{a(t)^3}
\label{eq:masscons}
\end{equation}
which is a conserved quantity at all times.  For reasons of
convenience, we also define the dimensionless variable
\begin{equation}
a_v(t)\equiv a_i\frac{r_v(t)}{r_v(t_i)}
\label{eq:aveq}`
\end{equation}
which, at very early times, gives $a_v(t)=a(t)$.  From equations
(\ref{eq:masscons},\ref{eq:aveq}) we get the relation
\begin{equation}
\frac{\dot{r}_v(t)}{r_v(t)}-
\frac{\dot{a}(t)}{a(t)}+\frac{\dot{\delta}_v(t)}{3(1+\delta_v(t))}=0 
\label{eq:deltareq}
\end{equation}

If the void is spherically symmetric, its evolution is described by
the Friedmann equation,
\begin{equation}
\left(\frac{\dot{r}_v(t)}{r_v(t)} \right)^2=\frac{8\pi G
\rho_v(t)}{3}+\frac{\Lambda_v}{3}-\frac{k_v}{R_v(t)^2}
\label{eq:fried1}
\end{equation}
where $\Lambda_v$, the cosmological constant within the void, is
identical to $\Lambda$ in the rest of the universe (thus the subscript
$v$ will be dropped from this term in subsequent discussion).  $R_v(t)$
is the (presently unknown) radius of curvature within the void, and
$k_v$ is the sign of the curvature within the void.

The right-hand side of equation (\ref{eq:fried1}) may be simplified
considerably.  Equation (\ref{eq:masscons}) yields the relation
\begin{equation}
\rho_v(t)=(1+\delta_v(t))\frac{\rho_0}{a(t)^3}
\end{equation}

From this, we may expand the right-hand side of equation
(\ref{eq:fried1}) as
\begin{eqnarray}
\label{eq:rhs1}
\frac{8\pi G \rho_v(t)}{3}+\frac{\Lambda}{3}-\frac{k_v}{R_v(t)^2}&=&
H_0^2 \frac{\Omega_M}{a^3}+H_0^2 \Omega_\Lambda+\delta_v(t)H_0^2
\frac{\Omega_M}{a^3}-\frac{k_v}{R_v(t)^2} \\ \nonumber &=&
\left(\frac{\dot{a}}{a}\right)^2+\delta_v(t)H_0^2
\frac{\Omega_M}{a^3}-\frac{k_v}{R_v(t)^2} \ ,
\end{eqnarray}
where we have substituted $(\dot{a}/a)^2$ for the right-hand side of
the Friedmann equation of the background cosmology.  Note that this
substitution assumes the background cosmology is spatially flat. It is
a trivial, albeit algebraically messy exercise to extend this analysis
to the case of non-flat background cosmologies.  Note also that in
pathologically closed universes, the void itself may be closed as
well.

We may also use equation (\ref{eq:deltareq}) to expand the left-hand
side of equation (\ref{eq:fried1}) as
\begin{eqnarray}
\label{eq:lhs1}
\left(\frac{\dot{r}_v(t)}{r_v(t)}
\right)^2&=&\left(\frac{\dot{a}}{a}\right)^2
\left(1-\frac{a\dot\delta_v(t)}{3\dot{a}(1+\delta_v(t))}\right)^2\\
\nonumber
&=& \left(\frac{\dot{a}}{a}\right)^2 \left(1-\frac{\delta_v(t)}{3}\right)^2 \\
\nonumber
&=& \left(\frac{\dot{a}}{a}\right)^2 \left(1-\frac{2 \delta_v(t)}{3}\right)
\end{eqnarray}
Prior to equation (\ref{eq:lhs1}), the relations we derive are fully
general. However, the second two equalities in equation
(\ref{eq:lhs1}) only hold at extremely early times in the evolution of
the void, since at early times $\delta_v \propto a$ and $|\delta_v| <<
1$.

Equations (\ref{eq:rhs1},\ref{eq:lhs1}) may be combined, yielding (at
early times only)
\begin{equation}
-\frac{2\delta_v(t)}{3}\left(\frac{\dot{a}}{a}\right)^2=
\delta_v(t)H_0^2
\frac{\Omega_M}{a^3}-\frac{k_v}{R_v(t)^2} 
\end{equation}
Substituting, once again, $H^2(t)=H_0^2(\Omega_M/a^3+\Omega_\Lambda)$,
we obtain
\begin{equation}
-\frac{5}{3}\delta_v(t)H_0^2 \frac{\Omega_M}{a^3}=\frac{1}{R_v(t)^2}\ ,
\end{equation}
where we have substituted $k_v=-1$ (necessarily the case for an
underdense void in a flat cosmology), or
\begin{equation}
R_v(t)=\frac{a}{H_0}\sqrt{-\frac{3a(t)}{5\delta_v(t)\Omega_M}}
\end{equation}
It is clear that $R_v\propto a_v$ at all times.  Thus, by
looking at the limiting case of $t\rightarrow 0$, we find at all times
\begin{equation}
R_v(t)=\frac{a_v(t)}{H_0}\sqrt{-\frac{3a_i}{5\delta_v(t_i)\Omega_M}}
\end{equation}
where the term in the radical is a constant provided $t_i$ is selected
sufficiently early. This ratio may be given as a fixed parameter of
the model.  As we will see, the ratio
\begin{equation}
\eta\equiv \frac{\delta_v(t_i)}{a_i}
\end{equation}
together with the parameters of the background cosmology
($\Omega_M,\Omega_\Lambda,H_0$) completely constrain the evolution of
the void region. We will compute the final underdensity of the region,
$\delta_v(t_0)$, the equivalent density parameter, $\Omega_{v,M}$, the
equivalent cosmological constant, $\Omega_{v,\Lambda}$, and the
equivalent Hubble constant, $H_{v}(t)$ as a function of the parameter
$\eta$.

Having specified $\eta$, we may again write the general form of the
Friedmann equation within the void (valid for all times)
\begin{eqnarray}
\label{eq:fried2}
H_v^2(t)=\left( \frac{\dot{a}_v(t)}{a_v(t)} \right)^2&=&
\frac{8\pi G \rho_v(t)}{3}+\frac{\Lambda}{3}-\frac{k_v}{R_v(t)^2} \\
\nonumber
&=&\frac{H_0^2 \Omega_M}{a_v^3}+H_0^2\Omega_\Lambda-H_0^2\frac{5\eta\Omega_M}{3 a_v^2}
\end{eqnarray}
To proceed, we need to solve a boundary value problem using
equation (\ref{eq:fried2}).  By integrating from $t=0$ to $t_0$, and
given the constraint that at early times, $a_v(t)=a(t)$, we solve for
the parameter
\begin{equation}
\alpha\equiv a_v(t_0)
\end{equation}
Because the void region expands faster than the background universe,
$\alpha > 1$.  Thus, $\alpha=\alpha(\eta)$, and we obtain the
relationships for the density contrast and effective cosmological
parameters in the void at the present epoch, in terms of the parameter
$\alpha$:
\begin{eqnarray}
\delta_v(t_0)&=&\frac{1-\alpha^3}{\alpha^3}
\label{eq:delta}\\
H_{v,0}&=&H_0
\sqrt{\frac{\Omega_M}{\alpha^3}+\Omega_{\Lambda}-\frac{5\eta\Omega_M}{3\alpha^2}}\\
\Omega_{v,M}&=&\frac{H_0^2 \Omega_M}{H_{v,0}^2\alpha^3} \ \ ; \ \ \
\Omega_{v,\Lambda}=\frac{H_0^2 \Omega_\Lambda}{H_{v,0}^2} \ \ ; \ \ \
\Omega_{v,k}=-\frac{5\eta\Omega_M H_0^2}{3 \alpha^2 H_{v,0}^2}
\label{eq:omega}
\end{eqnarray}
where $\Omega_k$ is the corresponding ``density'' parameter of the
curvature term in the Friedmann equation, such that $\Omega_M +
\Omega_{\Lambda}+\Omega_k=1$.

\begin{figure}
\centerline{\psfig{figure=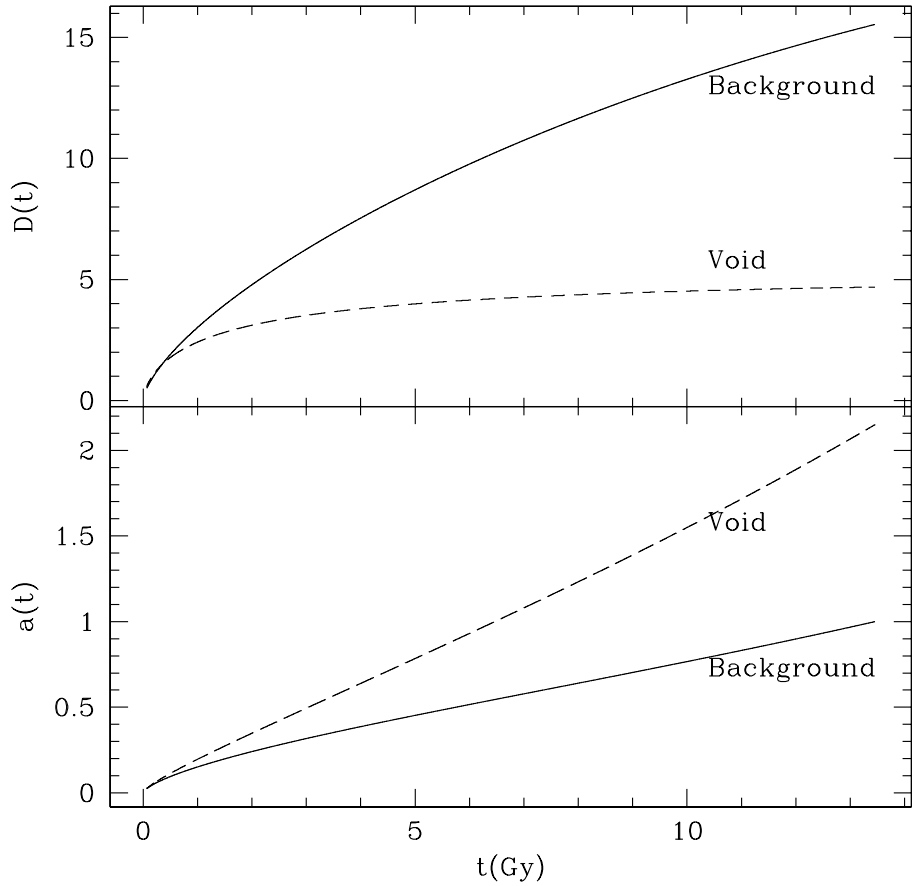,angle=0,height=5.0in}}
\caption{The growth of the scale-size and perturbations within a
  $\delta_0=-0.9$ void.  This assumes a background cosmology of
  $\Omega_M=0.3$, $\Omega_\Lambda=0.7$, and $h=0.7$.  Note that at
  early times the growth of structure within the void and in the
  background universe are identical but growth in the void is
  suppressed after $t\simeq 2Gy$ ($z\simeq 5$).}
\label{fig:growth}
\end{figure}

In the limit of $|\delta_0| << 1$, equation~(\ref{eq:delta}) reduces
to the linear growth relation discussed in, for example, Carroll,
Press \& Turner (1992; see also Lahav et al. 1991; Lightman \&
Schechter 1990),
\begin{equation}
\frac{\delta_{v}(t_0)}{\eta}\simeq
\frac{5}{2}\Omega_M\left[\Omega_M^{4/7}-\Omega_\Lambda+\left(1+\frac{1}{2}\Omega_M\right)\left(1+\frac{1}{70}\Omega_\Lambda\right)
\right]^{-1}\
\end{equation}
Figure~\ref{fig:growth} illustrates the different growth rates and
expansion factors for the void region and background cosmology.

However, voids may reach well into the $\delta_{v}(t_0)\simeq -0.9$
regime, thus the linear approximation is not valid and the void
universe must be treated numerically.  To accurately specify the
effective cosmological parameters for large voids, we numerically
integrate the Friedmann equation for a range of values of $\eta$. This
integration yields the parameter $\alpha$ and, from
equation(\ref{eq:delta}), the present-epoch void density contrast
$\delta_v(t_0)$.  By varying $\eta$ until the desired underdensity is
found, we can compute the effective parameters of the void cosmology.

\begin{figure}
\centerline{\psfig{figure=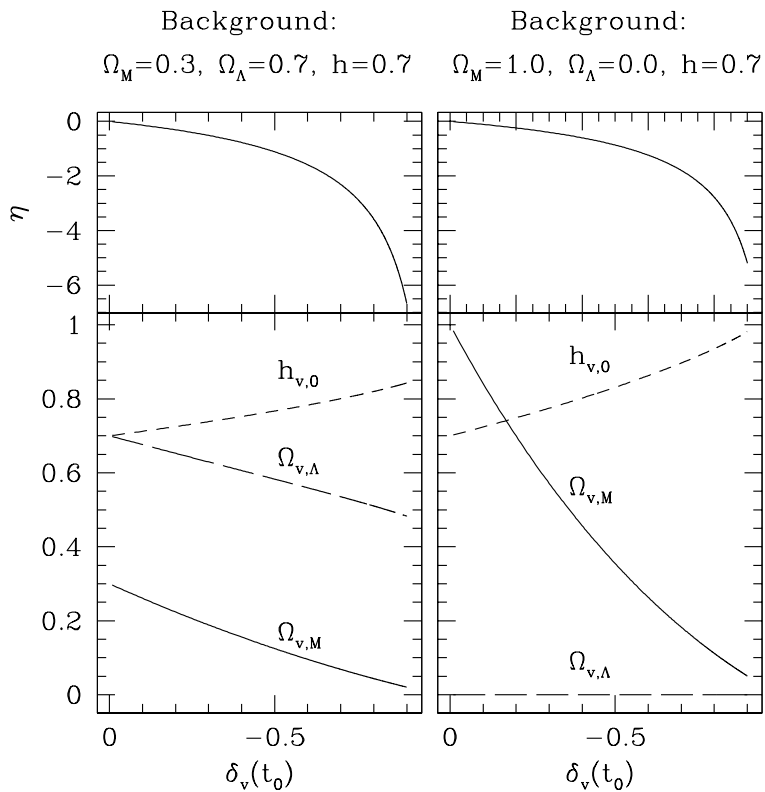,angle=0,height=5.0in}}
\caption{The corresponding cosmological parameters for the interior of
voids with varying overdensities, as described in
equations~(\ref{eq:delta}-\ref{eq:omega}).  All background
cosmological models are flat universes. The parameter $\eta$ is the
ratio of the void density contrast to the expansion factor at fixed
epoch, $\eta\equiv \delta_v/a$. The effective Hubble constant is
larger in voids because they expand relative to the background, while
the density parameters $\Omega_M$ and $\Omega_{\Lambda}$ are both
smaller (if non-zero).  }
\label{fig:params}
\end{figure}

In Figure~\ref{fig:params}, we show the relationships between the
density contrast of voids and the corresponding cosmological
parameters which would be needed to model a universe with the
appropriate age and dynamics. Deeper voids expand faster, thus they
have larger Hubble constant $H_v$. The matter density parameter
$\Omega_M$ is smaller, both because of the smaller average density and
because of the larger Hubble constant (recall the definition
$\Omega=(8\pi G \rho)/(3 H^2)$). While the cosmological constant
$\Lambda$ is identical in both the void and background, the
corresponding density parameter $\Omega_{\Lambda}$ is smaller in the
void because the void expands faster.
Because voids act like very open universes, the latter reflects the
phenomenon that dark energy is relatively less important in deep
voids, which become curvature-dominated quite early.  Even with a
large cosmological constant, structure formation in voids freezes out
at early time in similar fashion to a low-density  $\Lambda=0$
universe.

These results show that a simulation run with $h_v=0.84$,
$\Omega_{v,M}=0.021$, and $\Omega_{v,\Lambda}=0.48$ would dynamically
evolve just as the interior of a spherically-symmetric void with
$\delta=-0.9$ that resides within an $\Omega_M=0.3$ flat universe.
Obviously, the density parameters for the different components of the
non-relativistic matter ($\Omega_{DM}, \Omega_{baryon}$) should be
scaled relative to those of the background cosmology in the same way.
Any patch within such a simulation would mimic the
interior of a void, up to long-range tidal effects.  However, as we
discuss in the following section, we must also alter the input power
spectrum of fluctuations.

\section{The Primordial Power Spectrum in the Voids}
\label{sec:powerspec}

To correctly specify the input power spectrum for a simulation that
seeks to model the interior of large voids, we must make allowance for
two effects: First, the amplitude of the power spectrum of the density
contrast must be adjusted relative to the power spectrum for the
background cosmology, because the average matter density is smaller in
voids. Second, because we wish to simulate voids with fixed density
contrast, we should remove power on scales larger than the
characteristic scale of the voids themselves.  Here we discuss these
constraints on the initial power spectrum in running a simulation of
isolated void regions.  Again, the goal is to set up the initial
conditions so that any region within the simulation looks like the
inside of a large void, except for possible tidal effects near the
edges of the void.

\subsection{Small Scales}

Let us begin by discussing the power spectrum on scales smaller than
the void.  Consider that at the initial time, $t_i$, the density
fluctuations within some region can be given by
\begin{equation}
\delta(\vec{r})=\int d^3\vec{k} \hat\delta_{\vec{k}}
e^{i\vec{k}\cdot\vec{r}}\ ,
\end{equation}
where $\hat\delta_{\vec{k}}$ denotes the Fourier transform of the
density fluctuation field.  If the initial conditions are Gaussian, as
in the current standard models for structure formation, then each of
the Fourier components is statistically independent. Thus, the
contribution to the real-space density field from a particular
component is
\begin{equation}
\label{eq:fouriercomp}
d\rho(\vec{r})\propto\overline{\rho}\hat\delta_{\vec{k}}
e^{i\vec{k}\cdot\vec{r}}\ .
\end{equation}
On small scales, linear independence means that the contribution to
the density field from small scale modes is not dependent on whether
we are looking at a region which will evolve into a void or a cluster.
In void simulations of the type described above, $\bar{\rho}$ in
equation (\ref{eq:fouriercomp}) is the mean density in the void, not
the background universe.  Thus, we need to apply a correction to the
amplitude of each Fourier mode,
\begin{equation}
\hat\delta_{\vec{k}}(t_i) \rightarrow
\frac{\hat\delta_{\vec{k}}^{(BG)}(t_i)}{1+a_i \eta}\ ,
\end{equation}
where a superscript, $(BG)$, denotes a random density component in the
background cosmology.  Recall that $\eta\equiv \delta_v(t_i)/a_i$,
thus this correction is simply the ratio of the background density to
the average density within the void.  This correction yields the same
{\it mass} perturbations caused by small wavelength fluctuations as in
the background.  Note that in the limit of very high initial redshift
this correction almost completely disappears.

\subsection{Large Scales}

We also need to alter the input power spectrum on scales comparable to
and larger than the void scale, to tailor our simulations to mimic
only the interior of voids with specified density contrast.  Observed
large voids have density contrast $\delta\sim -0.9$ and typical radii
$R\sim 10h^{-1}$ Mpc.  If we consider that the entire simulation
volume is meant to be a ``void region'' then no power
should exist on scales much larger than $R$.  

The variance of fluctuations on a a scale $R$ is
\begin{equation}
\sigma^2(R,t_i)=\int 4\pi k^2 dk P^{(BG)}(k,t_i) |\hat{W}(kR)|^2
\end{equation}
where $\hat{W}(kR)$ is the Fourier transform of the spatial window
$W_R(x)$ over which the fluctuations are sampled (e.g., a spherical
tophat or Gaussian window) and $P^{(BG)}(k)$ is the background power spectrum
of fluctuations, $P(k)^{(BG)}=\langle|\hat{\delta}(k)|^2 \rangle$.
The effect of modifying the amplitude of the Fourier components 
is to modify the power spectrum as
\begin{equation}
P(k)=f(k)P^{(BG)}(k)
\end{equation}
where $f(k)$ is the squared modulus of the factor that multiplies each
Fourier component $\hat{\delta}(k)$.  On scales comparable to and
larger than the void, we want to smooth away the very fluctuations
that led to the creation of a void with density contrast $\delta_i$,
such that the amplitude of {\it suppressed} power is
\begin{equation}
\sigma^2_{SUP}(R,t_i)=\int 4\pi k^2 dk [1-f(k)]P^{(BG)}(k,t_i)  |\hat{W}(kR)|^2
\simeq \delta_i^2
\label{eq:sigmaout}
\end{equation}
This suppresses sufficient power that the rms of the suppressed
structure is similar to the mean underdensity of the void region.

For example, applying a Gaussian cutoff to large-scale power implies
\begin{equation}
f(k)=1-\exp[-(k/k_0)^2]
\end{equation}
as the smoothing function (where $k_0$ is the smoothing scale,
selected to satisfy equation (\ref{eq:sigmaout}).  The result will be
little structure in the simulations at wavenumber $k<k_0$.  Since we
do not wish to run a serious of identically periodic voids, this
implies that we should limit the size of the simulation box to $L\sim
2\pi/k_0$.  Of course, limiting $L$ effectively removes power on large
scales.  Note that the cutoff scale $k_0$ corresponds to a smaller
comoving scale than the current void diameter, because all the wave
modes within the void grow with time relative to the background.

From above, the smoothing function should approach
$f(k)=1/(1-a_i\eta)^2$ on small scales. 
Together, the large and small-scale constraints yield the modified
initial power spectrum for our void simulations:
\begin{equation}
P(k,t_i)=P^{(BG)}(k,t_i)\left[\frac{1-\exp[-(k/k_0)^2]}{(1-a_i\eta)^2}\right]
\end{equation}

\section{A Recipe for Cooking Voids}
\label{sec:recipe}

Suppose we wish to simulate a void with density contrast
$\delta_v(t_0)$ (which specifies the parameter $\alpha$) and comoving
diameter at $z=0$ of $L_0$ (in units of $h^{-1}$Mpc, where $h$ is the
dimensionless Hubble parameter of the background cosmology).  
There
are three separate steps that require careful accounting for the units
of distance and scale factors in the void and background cosmologies:
(1) generating initial density and velocity fields; (2) setting
parameters of the simulation code to evolve those fields; and (3)
translating void simulation results back into comoving coordinates of
the background cosmology (i.e., back into {\it real} comoving
coordinates). Here we review these steps.

To generate the initial density and velocity fields, note that the
fluctuations in the void region are those of a patch that has comoving
size $L_0/\alpha$.  This is smaller than the $z=0$ comoving size of
the void region because the void has expanded relative to the
background, thereby stretching all the wavemodes within it.  Thus, the
amplitude of the fundamental mode of the simulation cube should be set
by $P(\tilde{k})$, where $\tilde{k}=2\pi\alpha/L_0$.  The initial
power spectrum must be generated using the {\it background}
cosmological parameters, with amplitude correct for the specified
initial redshift $z_i$, but with shape and amplitude modified by the
factor $f(k)$ as discussed above. Scaling the velocities, both in the
initial conditions and in the outputs is very code-specific, as it
depends on the units.

The input parameters for running the simulation include the void
cosmological parameters, as well as the comoving box size, initial
redshift, and final redshift.  If specified in units of $h_v^{-1}$Mpc,
the comoving box size should be set to $L_v=L_0 (h_v/h)$.  The initial
redshift of the void simulation, $z_{v,i}$, must be larger than the
true $z_i$, by $(1+z_{v,i})/(1+z_i)=\alpha$. 

To correctly interpret the results of the simulation, we must account
for the difference in scale factors and Hubble constants in the
background cosmology and the void simulation. This is because,
although the void expands in comoving coordinates when embedded in the
background, it does not, of course, grow in comoving coordinates when
it sits inside the effective void cosmology.  At $z=0$, scales in the
void simulation are exact when expressed in Mpc (not in
$h_v^{-1}$Mpc).  Therefore, to get scales in the typical units of
$h^{-1}$Mpc, we must multiply by the ratio of Hubble constants, $L=L_v
(h/h_v)$. At any other epoch, we must also account for the different
growth factors, thus the general expression is
\begin{equation}
L=L_V{h\over h_v}{a_v(t)\over \alpha}
\end{equation}
 (recall that, by
definition, $a_v(t_0)=\alpha$).  At the initial redshift, $z=z_i$,
$L=L(h/h_v)/\alpha$, as discussed above.

\section{A Mass-Function Sanity Check}
\label{sec:massfunc}

In developing the formalism for a void-only simulation, we have made a
number of simplifying assumptions, and objections may be raised as to
whether the assumptions of spherical symmetry (demanded by Birkhoff's
theorem) or fixed total mass (demanded by running the void in
isolation) are well-founded.  

\begin{figure}
\centerline{\psfig{figure=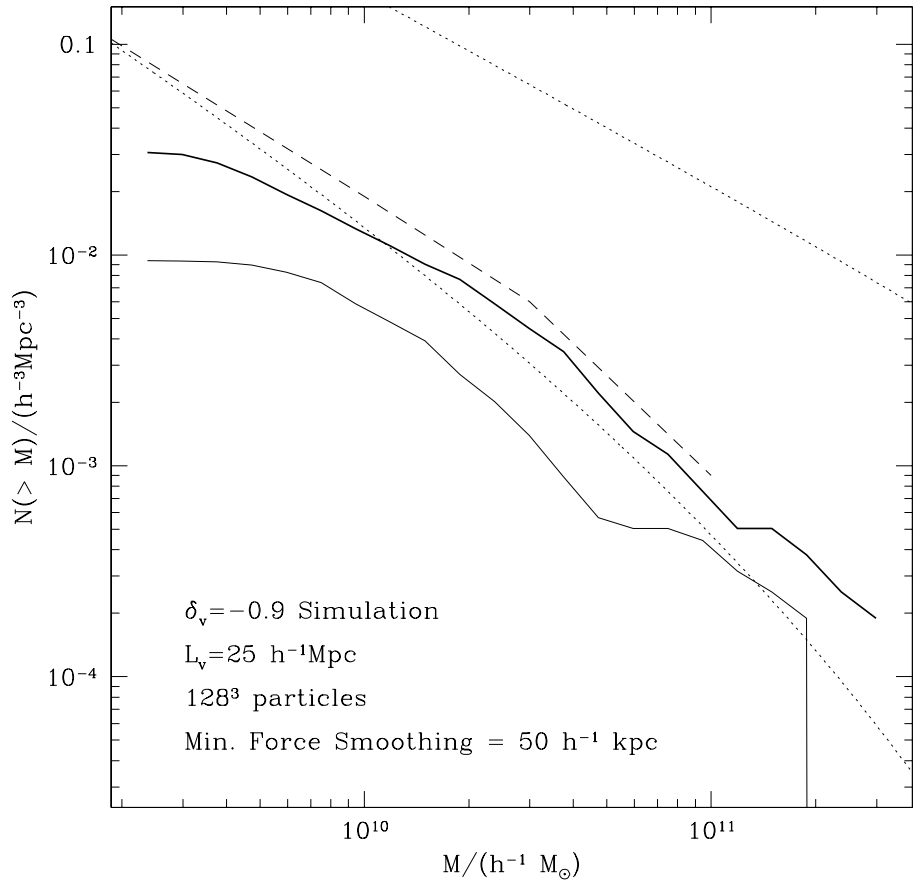,angle=0,height=5.0in}}
\caption{The dark matter halo mass function of a $128^3$ particle,
$\delta_v=-0.9$ void-only simulation, as described in the text (bold
solid).  The dashed line shows the mass function computed in the void
regions of the Gottl\"{o}ber et al. (2003) simulations, while the
dotted lines show the theoretical mass functions of $\delta_v=-0.5$
(top), and $\delta_v=-0.9$ void regions using Press-Schechter
analysis.  The thin solid line represents the same simulation with no
force refinement, which produces a notably worse fit to both other
simulations, and to theory.}
\label{fg:mf}
\end{figure}

As a ``sanity check,'' we have run several relatively modest $128^3$
particle simulations of a $\delta_v=-0.9$ void region.  We used Enzo
(Norman \& Bryan 1999), an adaptive mesh refinement (AMR)
hydrodynamics code that includes a grid that can adapt automatically
to provide high-resolution in regions of interest. Refinement was set
to allow resolution down to $1/512$ the simulation box-size (25
$h^{-1}$ Mpc), approximately 50 $h^{-1}$ kpc, and only gravitational
forces were used.  We then used the HOP (Eisenstein \& Hut 1998)
group-finder to identify galaxies, and computed the mass-function of
the dark matter halos in the void region.  The results of this simulation can
be found in Figure~\ref{fg:mf}.

We compare the void-only simulation mass function with the mass
function found by Gottl\"{o}ber et al. (2003), who simulated a large
region of the universe, excised the void galaxies and computed their
mass function. For consistency, the results cited are based on their
$\delta_v < -0.9$ sample.  We also compute a theoretical mass function
based on Press-Schechter (1974) analysis of the spherical collapse of
halos.  The theoretical curves plotted represent voids of
$\delta_v=-0.5$ (top), and $\delta_v=-0.9$ (bottom) using the analytic
form of Goldberg et al. (2004).  The form of Sheth \& Tormen (2002)
produces a nearly identical curve, but differs in assumption, since
the Goldberg et al. result assumes the fully nonlinear (semi-analytic)
density contrast of the void, derived herein, and uses the size of the
void as an additional Bayesian prior.

It is clear that at most masses, the void-only simulation produces an
excellent fit with both larger simulations and purely analytic
results.  The slight discrepancy at the low-mass end can be attributed
to limited force resolution in the present simulations. As seen in
Fig.~\ref{fg:mf} a simulation with no force refinement produces a
significantly worse fit of the low-mass mass function.

\section{Discussion and Future Prospects}
\label{sec:future}

To summarize, the
following are the steps to run a void simulation using our new
approach: (1) Select the present-epoch underdensity $\delta_v(t_0)$ of
the voids of interest. (2) Select the parameters of the background
cosmology in which the voids reside. (3) Compute the effective
cosmological parameters ($\Omega_{v, M}$, $\Omega_{v, \Lambda}$,
$H_v$) of the interior of the void, following Section
\ref{sec:parameters} (this requires numerical integration of the
Friedmann equation of the void to compute the appropriate value of
$\alpha$). (4) Compute the
modified initial power spectrum of fluctuations, following Section
\ref{sec:powerspec}. (5) Run a simulation using these parameters and
power spectrum, in a box with comoving size $L$, such that the box
size is not a large multiple of the void diameter. (6) Rescale the
simulation results to adjust for the difference in comoving scales in
the voids-only simulation and the background cosmology.

Using only a few extra steps in preparing the simulation, we
significantly improve the efficiency of our simulations. For fixed
computational resources. we gain at least an order of magnitude in
mass resolution: The voids of interest fill at least $\sim 40\%$ of
the volume of the universe but have $\sim 10\%$ the mean density of
galaxies, thus a voids-only simulation can use the same number of
particles to simulate the formation of a factor of $\sim 25$ fewer
galaxies at a fixed level of force refinement (because we simulate
only the void galaxies of interest). High-level adaptive simulations
of larger regions will, at best, asymptotically approach the
efficiency of the proposed method.  Simulating only the void region
also affords a small increase in spatial resolution.  This approach is
in contrast to running a large simulation, then excising the
underdense regions for closer inspection.  In the latter, huge
computational resources are expended on regions which are
uninteresting for the analysis of voids, i.e., clusters, which soak up
lots of computational cycles.

This approach for running specialized void simulations is valid for
studying the interior of large voids.  We intentionally smooth out
structure on the largest scales. Thus, any contribution of tidal
structure to the internal dynamics of voids cannot be studied by the
current approach.  Nor, for that matter, can the large scale internal
structure of voids be examined, as the edges are likely to be strongly
affected by nearby structure.

The next step in this analysis is to apply this approach to run large
hydrodynamic simulations of void regions.  We will use this technique
to make projections of the mass spectrum of galaxies in voids
(Goldberg et al. 2004), as well as their photometric, morphological,
and spectroscopic properties.  These simulations will be used to
interpret results from analysis of large samples of void galaxies
being identified from the Sloan Digital Sky Survey (Rojas et
al. 2003a,b; Hoyle et al. 2003), as well as to guide analysis of
future deeper surveys.

\acknowledgements

DMG acknowledges support from NSF grant AST-0205080.  MSV acknowledges
support from NSF grant AST-0071201 and a grant from the John Templeton
Foundation. We thank Fiona Hoyle and Henry Winterbottom for useful
conversations, and Greg Bryan for use of the Enzo code and significant
guidance in its use.  We thank the anonymous referee for several
helpful suggestions to the final draft.

\end{document}